\input harvmac.tex
\Title{\vbox{\baselineskip14pt\hbox{HUTP-97/A066}\hbox{hep-th/9711067}}}
{Black Holes and Calabi-Yau Threefolds}
\bigskip\vskip2ex
\centerline{Cumrun Vafa}
\vskip2ex
\centerline{\it  Lyman Laboratory of Physics, Harvard
University}
\centerline{\it Cambridge, MA 02138, USA}
\vskip .3in
We compute the microscopic entropy of certain 4 and 5 dimensional
extermal black
holes which arise for compactification of M-theory and type IIA
on Calabi-Yau 3-folds.  The results agree with macroscopic predictions,
including some subleading terms. The macroscopic entropy in the 5 dimensional
case predicts a surprising growth in the
cohomology of moduli space of holomorphic curves in Calabi-Yau threefolds
which we verify in the case of elliptic threefolds.

\Date{\it {Nov. 1997}}
%\draft

\newsec{Introduction}
Microscopic degeneracy counts of BPS states in
4 and 5 dimensional compactifications and agreement
with the corresponding Bekenstein-Hawking formula for
black hole entropy has been tested in many cases.
In this paper we extend these results by considering
a wide class of supersymmetric
compactification of M-theory to 4 and 5 dimensions.  In particular
we consider compactifications of M-theory on elliptic
Calabi-Yau threefolds to 5 dimensions, and type IIA compactifications
on Calabi-Yau threefolds to 4 dimensions and verify the match with the
macroscopic prediction of Bekenstein-Hawking Entropy formula. The method
we use is unaffected whether the Calabi-Yau threefold has $SU(3)$
holonomy, $SU(2)$ holonomy (such as $K3\times T^2$) or trivial holonomy
( $T^6$).  When we refer to Calabi-Yau threefolds in this paper,
we have any of these three possible cases in mind (in particular
our work includes as a special case the results obtained in
\ref\sv{
A. Strominger and C. Vafa, Phys. Lett. {\bf B379} (1996) 99.}).

  In the case of 5-dimensional BPS states we perform the counting
in two different ways, one using properties of F-theory  and its
relation to M-theory, and the other
using a direct count of M2 branes wrapped around 2-cycles of CY 3-folds.
The latter is not only consistent with computations
of holomorphic curves in CY 3-folds based on mirror symmetry
but actually sheds light on some aspects of it.
For the case of 4-dimensional compactifications of
type IIA strings, in a recent
paper
\ref\msw{J. Maldacena, A. Strominger and E. Witten, hep-th/9711053.}\
it was shown how
properties of  M-theory 5-brane and its relation
to type IIA branes leads to a
prediction of BPS states in agreement
with macroscopic results. Here we show that the same results follow from
a direct type IIA
description as well in the spirit of BPS count of type IIA
strings on $K3$ and $T^4$ \ref\gas{C. Vafa,
Nucl. Phys. {\bf B463} (1996) 415\semi Nucl. Phys.
{\bf B463} (1996) 435.}\ref\bsv{M. Bershadsky,
V. Sadov and C. Vafa, Nucl. Phys. {\bf B463} (1996) 420.}.
Thus the situation in 4 dimensions is
 completely parallel with the 5 dimensional case
which can also be computed in two different ways.

\newsec{F-theory, M-theory, Type IIA Chain}
Let us review some aspects of F-theory, M-theory and Type IIA theories
which are relevant for us.  In particular the relation between
F-theory and M-theory will prove useful for understanding the
entropy of 5d black holes
and the relation between M-theory and type IIA will prove useful
for the computation of entropy of 4d black holes.
\subsec{F-theory, M-theory duality}
Consider F-theory on an elliptic manifold $K$ with a section
$B$, which is the base of the elliptic fibration.
Moreover consider a three brane of type IIB wrapped around
a holomorphic two cycle $C\subset B$ which will correspond to a string in
the uncompactified space time.
Upon further compactification of F-theory on circle, we
obtain M-theory on the same manifold $K$ \ref\ef{C. Vafa,
Nucl. Phys. {\bf B469} (1996) 403.}.
Moreover one can map what corresponds to the string
obtained by wrapping F-theory 3-brane around $C$ in the M-theory
setup.  There are two
possibilities
to consider upon further compactification on a circle: A string which
is wrapped around the circle, carrying some momentum $p=n/R$ along
the circle, or a string which is unwrapped.  In terms of M-theory
compactification on $K$, the wrapped strings with momentum $n$ correspond to
$M2$ branes wrapped around a 2-cycle in $K$ in the homology class
$[C]+n[E]$ where $[E]$ denotes the class of the elliptic fiber.
The strings which are unwrapped correspond to $M5$ branes wrapped around
the four manifold $\widehat{C}$ which is the total space of the curve $C$
together
with the elliptic fiber on top of it.  This is quite an interesting link:
If we wish to count the number of M2 BPS states wrapped in the class
$[C]+n[E]$, it suffices to compute the number of BPS states of a wrapped
string of momentum $n$.
The easiest way to find the degrees of freedom on this string is
to consider a further compactification on $S^1$ and consider not
wrapping the string around $S^1$ and use the M-theory
description of the string.    The reason for doing this, instead of
a direct count of the low energy modes on the string in the
6-dimensional setup is that the coupling constant on the 3-brane worldvolume
is undergoing $SL(2,{\bf Z})$ monodromy and  it is thus
difficult to formulate a Lagrangian description of
the low energy degrees of freedom on the left-over string.
The degrees of freedom on this string, apart from compactification of one
bosonic direction of the F-theory string, is
the same degrees of freedom as those of the string obtained by wrapping
$M5$ brane on $\widehat C$.  Let us count these degrees of freedom.
The string we obtain by wrapping M5 brane on $\widehat C$ has
$(0,4)$ supersymmetry.  To compute the number of BPS states
it suffices to compute the left-moving degrees of freedom
on the string.  Apart form the center of mass degree of freedom,
the left-moving bosonic modes arise from the modes corresponding
to moving the $\widehat C$ in the Calabi-Yau, which has complex
dimension $h^{2,0}({\widehat C})$ as well as anti-self dual
harmonic two forms on $\widehat C$ which correspond
to left-moving degrees of freedom on the string due to self-duality
of the anti-symmetric field strength on $M5$ brane.  The number
of anti-self dual
2-forms is equal to $h^{1,1}({\widehat C})-1$.  Similarly the left-moving
fermions on the string come from the odd cohomologies of $\widehat C$,
which is equal to $4 h^{1,0}({\widehat C})$ real fermions.  In addition
there are three scalar modes corresponding to the center of mass
of the string.  So all told the number of real left-moving
bosons is
$$2h^{2,0}+(h^{1,1}-1)+3=h^{0,0}+(h^{2,0}+h^{1,1}+h^{0,2})+h^{2,2}=b^{even}(
{\widehat C})$$
where $b^{even}$ denotes the total number of even cohomology
elements and
 we have used $h^{0,0}=h^{2,2}=1$ and $h^{2,0}=h^{0,2}$.  Moreover
the number of left-moving fermions is equal to
$$h^{1,0}+h^{0,1}+h^{2,1}+h^{1,2}=b^{odd}({\widehat C})$$
where $b^{odd}$ denotes the total number
of odd cohomology elements.
The number of BPS states of this string with left-momentum
$n$ and no right-moving momentum would correspond to the
coefficient of $d(n)$ in the expansion
$$\sum d(n)q^n={[\prod_n (1+q^n)]^{b^{odd}}\over [\prod_n
(1-q^n)]^{b^{even}}}$$
Using Hardy-Ramanujan formula it is easy to see that for large
enough $n$ we have
\eqn\gro{d(n)\sim {\rm exp}(2\pi\sqrt {n c_L\over 6})={\rm exp}(
2\pi\sqrt
{n(b^{even}+{1\over 2} b^{odd})\over 6}}

Now let us compare this to the direct count of $M2$ branes wrapped
in the class $[C]+n[E]$. The mathematical problem involved
turns out to be simpler if we consider one compact spatial direction
in which case we obtain type IIA on the same manifold, and instead
of wrapped M2 branes we consider wrapped D2 branes\foot{
The same comment applies to counting the BPS degeneracies
of wrapped M2 branes of M-theory on K3.  The direct
count for M2 branes in the $K3$ case and the present
case is also possible \ref\sava{V. Sadov and C. Vafa, work in progress.}.}.
  Clearly
the physics of the counting will not change as we are not
considering any excitations along the compact direction, except
for a possible quantized momentum along the
compact direction.  We will show that the counting of the states
in the Type IIA setup will give the same result for all values
of discrete momentum along the extra compactified
circle, and thus in the large
radius limit (strong coupling
limit of type IIA) will give the BPS spectrum for a
particle in one higher dimension. The number of BPS
states for type IIA in this setup case can be computed
by using similar techniques as was done in \bsv
(see also the review article \ref\va{C. Vafa,
``Lectures on Strings and Dualities," hep-th/9702201.}). The relevant D2
brane configuration consists of bound states of $n$ D2 branes
wrapped around $[E]$ with one single $D2$ brane $[C]$. The moduli
space of such bound states is the same as the choice of $n$
points on $[C]$ to which the $D2$ branes wrapped around $E$
are attached. In other words, the total configuration of the
$D2$ brane looks like a singular Riemann surface, consisting of
$C$ with $n$ ``elliptic ears''.  If we include the choice
of Wilson line on the elliptic D2 brane, which is equivalent for
each torus to the choice of a point on the dual torus, we see
that the full moduli space involves the choice of n points on the
$\widehat C$ manifold, where the elliptic fiber is the dual
to the original one. Of course the order of the points is irrelevant and
  thus the number of BPS states is the same
as
\eqn\eulc{H^*(Sym^n({\widehat C}))}
where $H^*$ denotes the cohomologies and $Sym^n$
denotes the $n-$fold symmetric product. Note that if we turn
on $m$-units of $U(1)$ flux on the D2 brane, which would correspond
to considering momentum $m/L$ along the extra compact dimension,
the moduli space we obtained above would not change as that
would still give the Jacobian as the moduli space.  We thus see
the count of BPS states will give the same result for arbitrary momentum
$m/L$ and so in the large $L$ limit (strong coupling limit of type IIA)
we obtain an M2 prediction.

We are thus left to compute the cohomologies \eulc .
  This question
was encountered before in the course of studying $N=4$
gauge theories in 4-dimensions \ref\vw{C. Vafa and E. Witten,
Nucl. Phys. {\bf B431} (1994) 3.}\ and also
in the math literature \ref\matli{L. Gottsche, Math.
Ann. {\bf 286} (1990) 193.}\ where one uses
orbifold techniques to compute it.  One finds that the cohomology
can be computed by considering the Fock space of $b^{even} $ bosonic
oscillators and $b^{odd}$ fermionic oscillators, exactly
in accordance with what we found in the context of wrapped
string of F-theory.  Thus they agree, as anticipated by duality
between F-theory and M-theory.\foot{The similarity of this
result with recent results \ref\egu{T. Eguchi, K. Hori and
C.-S. Xiong, Phys. Lett. {\bf B402} (1997) 71\semi
T. Eguchi, M.
Jinzenji and C.-S. Xiong, ``Quantum Cohomology and
Free Field Representation," hep-th/9709152.}\ref\kat{S. Katz, unpublished.}\
showing the existence of a
Virasoro action on partition function of topological strings which
count holomorphic curves is very suggestive.}  This also explains
part of the results in \ref\kmv{A. Klemm, P. Mayr and C. Vafa,
``BPS States of Exceptional Non-Critical Strings,''
hep-th/9607139.}\ref\gan{O. J. Ganor, Nucl. Phys. {\bf B479} (1996) 197.}\
where it was observed that the $E_8$ string carries
12 bosonic oscillators on the worldsheet--12 is the number
of even cohomologies of $P^2$ blown up at 9 points, which
gives rise to $E_8$ string by wrapping the M5 brane around it.

\subsec{M-theory, type IIA duality}
If we consider M-theory on an arbitrary CY 3-fold ${\cal K}$
(not necessarily elliptic), further compactification on $S^1$ gives
a type IIA theory on manifold ${\cal K}$ \ref\ptow{P. Townsend,
Phys. Lett. {\bf B350} (1995) 184.}\ref\ew{
E. Witten, Nucl. Phys. {\bf B443} (1995) 85.}.
  Now consider a holomorphic four-cycle
${\cal C}$ in ${\cal K}$.  Then if we consider the M5 brane
wrapped around ${\cal C}$ with some left-moving momentum $n/R$
along the extra circle, the computation is identical to what we did
above and was recently considered in \msw .
  From the type IIA perspective this has the interpretation
of a $D4$ brane wrapping around ${\cal C}$ bound to $n$ 0-branes
(modulo the shift induced by wrapped $D4$ brane which is $-c_2({\cal C}/24)$
\bsv \ref\ghm{M. Green, J.A. Harvey and G. Moore, Class. Quant. Grav.
{\bf 14} (1997) 47.}), as is
standard from comparing the branes of M-theory with those of type IIA.
However this count of BPS states can also be done quite directly
in the type IIA setup as was
done in
\va : Namely each 0-brane binds to the 4-brane ${\cal C}$ \ref\sen{
A. Sen, Phys. Rev. {\bf D54} (1996) 2964\semi Phys. Rev.
{\bf D53} (1996) 2874.}.  Thus
the moduli space of this bound state involves the choice of $n$ points stuck
on ${\cal C}$ and we thus obtain again the moduli space of symmetric product
of $n$ points on ${\cal C}$ as the moduli space of bound states,
whose cohomology again gives the same answer as mentioned before.  In fact
the agreement between the 5-brane count of BPS states \ref\dvv{R.
Dijkgraaf, E.
Verlinde and H. Verlinde, Nucl. Phys. {\bf B486} (1997) 89.}\
and the type IIA
count \va \bsv\ in the special cases of $K3$ and $T^4$ was already known.
Here we have shown this to be true for more general four manifolds.

\newsec{Black holes in 5 dimensions}
Consider M-theory compactified on a Calabi-Yau 3-fold $K$
 down to 5 dimension.  Let
$Q$ denote an element of $H_2(K,Z)$
(corresponding to the M2 charge).  Let $k$
denote the Kahler class of Calabi-Yau.  Then the BPS mass
of this state is given by
$$M=\int_{[Q]} k$$
The entropy predicted for this configuration according to the results
\ref\kall{S. Ferrara and R. Kallosh, Phys. Rev. {\bf D54} (1996) 1525
\semi 1514.}\ is
\eqn\entf{S=a{\rm Min}_{k} [ M^{3/2}],\qquad {{\rm subject}
\quad {\rm to} \qquad \int_K k^3=1}}
where $a$ is a universal constant.  In other words
we vary the kahler
class $k$, subject to the total volume of
the CY being 1 and then compute the mass $M$ for this particular
$k$, in terms of which the entropy formula
has the universal form given above. For the application
we have in mind we will assume that $K$ is elliptic CY 3-fold
with a section $B$.  Let $C$ denote a curve (i.e. a Riemann
surface) in $B$.  Let $e_i$ denote a basis for $H_2(B,Z)$.
Let us define
$$k_E=\int_{T^2} k \qquad k_i=\int_{e_i} k \qquad d_{ij}=e_i\cdot e_j\qquad
[C]=C^ie_i$$
Let us consider the charge state
$$Q=n[T^2]+[C]$$
Then
$$M=nk_E+C^ik_i$$
which is to be minimized as we vary $k_E$ and $k_i$ subject to $V=k_E
k_id^{ij}k_j=1$. This can be easily done and one finds that at the minimum
$$M={3\over 4}( n C\cdot C)^{1/3}$$
Moreover one finds that
at the minimum $k_E=4^{-1/3}n^{-2/3}(C\cdot C)^{1/3}$
and $k_i=4^{1/6}C_i n^{1/3} (C\cdot C)^{-2/3}$).  Using the universal
entropy formula (including the universal numerical
factor) this gives
$$S=2\pi\sqrt{{1\over 2} C\cdot C n}$$
Now we come to the microscopic computation.  From the discussion
in the previous section it is clear that all we need to do is to compute
the cohomology of $\widehat C$, which is the four manifold
consisting of the elliptic manifold over the Riemann surface $C$.
It is straight forward to do this.  Let $c_1$ denote the first Chern
class of the base $B$.  Then one can show using index theory
and the vanishing of the first chern class of $K$ that
$$h^{2,0}({\widehat C})={1\over 2}(C\cdot C+c_1(C))$$
$$h^{1,0}({\widehat C})={1\over 2}(C\cdot C-c_1(C))+1$$
$$h^{1,1}({\widehat C})=C\cdot C+9 c_1(C)+2$$

We thus see that
$$b^{even}+{1\over 2}b^{odd}=(3 C\cdot C+9 c_1(C)+6)$$
Thus according to \gro\ the Hardy-Ramanujan formula gives that for large $n$
$$S=2\pi \sqrt{{1\over 6} n(3 C\cdot C +9 c_1(C)+6)}$$
To leading order in large $n$ and $C$, this is in agreement
with the black hole formula above.  The microscopic prediction
has some subleading terms which do not vanish for large charges.
The leading correction goes as
$${3\pi }\sqrt{n c_1(C)^2\over 2 C\cdot C}$$
This is similar to what has been found in certain 4d cases
using M-theory recently
\msw\ as we will rederive below in the context of type IIA.
In that case the subleading terms were matched with corrections
to the effective action in 4 dimensions involving $R^2$ type terms
which are proportional to Euler characteristic in 4 dimensions.
In fact there are such corrections already in the 5-dimensional
theory \ref\fera{I. Antoniadis, S. Ferrara,
R. Minasian and K.S. Narain,``$R^4$ Couplings in M and Type II Theories,'' hep-th/9707013.}
which go as
 $$[\int_K c_2(K)\wedge k] R^2$$
Here the $R^2$ is not the Euler characteristic
anymore (as that vanishes in 5 dimensions) but rather
it corresponds to a specific contraction of two R's
which in 4d would give the Euler character density.
If we consider a Euclidean black hole, this term corrects
the entropy by ${\rm exp}(-\delta S)$.
We will now verify that the leading correction this induces
has the same charge dependence as was found by the microscopic
derivation above.

For large
$n$ the dominant term comes from the expansion involving
$k_i$ of the base as can be
seen from the fixed point values of moduli
at the horizon given above.  Let $\widehat e_i$ denote the
elliptic four manifold over $e_i$ which form a basis
for $H_2(B)$.  Then $c_2({\widehat
e_i})=12c_1(e_i)$, where $c_1$ is the first chern class of the base $B$.
Expanding the effective action the leading term will thus come from
$$R^2 c_1(e_i)k_i$$
Integrating $\int R^2$ over the 5-dimensional extremal black hole
will give a term which scales as (length) (as $R^2$ has dimension
4).  Since the mass in the
extremal black hole appears in the combination
$M/r^2$, this means that the integral will go as $M^{1/2}$.
Using the fact that $M\sim [n (C\cdot C)]^{1/3}$ and that $k_i\sim
C_i n^{1/3}(C\cdot C)^{-2/3}$
we learn that the correction to the effective action in the presence
of the 5d black hole
scales as
$$[n(C\cdot C)]^{1/6}(c_1(e_i)C_i n^{1/3}(C\cdot C)^{-2/3})=
\sqrt{ n c_1(C)^2\over C\cdot C}$$
which has precisely the same charge dependence as predicted
by the subleading corrections from the microscopic computation.
It would be interesting also to check that the numerical
coefficients in front also matches--this would require computing
the integral of $R^2$ over the 5d extremal euclidean black hole,
which we have not performed.

In this section we have mainly considered an elliptic Calabi-Yau
in order to do the explicit count of holomorphic curves.
The question arises as to whether we can do this count more generally
for M-theory compactifications on
 arbitrary Calabi-Yau manifolds.  In this case
the following puzzle has been pointed out by Strominger \ref\ast{
A. Strominger, private communication.}:  The black hole entropy
formula predicts that the entropy of holomorphic
curves should scale as the $N^{3/2}$
as we rescale the $M2$ charges $Q\in H_2({\cal K})$ by
$Q\rightarrow N Q$.  However, in some examples, such
as quintic it is known that the entropy of genus zero curves
scales as $N$ (i.e. the number of rational curves
of degree $N$ grows as ${\rm exp}(aN)$ for some $a$)
\ref\cand{P. Candelas, X. De La Ossa, P.S. Green and L. Parkes,
Nucl. Phys. {\bf B359} (1991) 21.}.
One could hope to overcome this by assuming that the higher
genus holomorphic curves will grow at a different rate.  However
this hope is dashed by the results in \ref\bcov{M. Bershadsky,
S. Cecotti, H. Ooguri and C. Vafa, Nucl. Phys. {\bf B405}
(1993) 279.}\ where it was
shown that at least for the quintic 3-fold the growth in entropy is again
linear in $N$.  Moreover it was conjectured in \bcov\ based
on universality arguments (and relations with non-critical
$c=1$ strings)
 that the entropy being linear in $N$
should be a generic behavior for generic Calabi-Yau threefolds.
We seem to have arrived at a serious contradiction!  This is not so
and we will now explain the resolution of the puzzle.

The computation done by mirror symmetry is not strictly speaking
`counting' holomorphic curves, but rather computing
the Euler characteristic of the moduli space
\foot{More precisely the Euler characteristic
of a specific bundle over the moduli space.}.  For example
if we consider $T^6$ the mirror symmetry answer for
the `count' of holomorphic
curves gives zero, which should be interpreted as the statement
about the Euler characteristic of the moduli space of such curves.
In fact clearly in the case of $T^6$ the construction above
 indicates that there is a huge growth in the
number of cohomologies of the moduli of holomorphic curves.
  Also
there are examples known where the `number' of curves is
negative \ref\ya{S. Hosono, A. Klemm, S. Theisen and S.-T. Yau,
Nucl. Phys. {\bf B433}( 1995) 501.}\ref\ket{P. Candelas,
A. Font, S. Katz and D.R. Morrison, Nucl. Phys. {\bf B429} (1994) 626.}
implying that there is a moduli space of curves with more
odd classes than even classes.  Thus the only resolution of the puzzle
for the discrepancy
between the microscopic and macroscopic prediction of black hole entropy
will be if for large enough $N$ we {\it always} get a non-trivial moduli
space of holomorphic curves\foot{
This is for cases of directions in $H_2$ for which
there is a finite horizon black hole, which exists if there
exists a kahler class which squares to the particular element in $H_2$.},
 and moreover the number
of even and odd cohomologies each go to leading order
as ${\rm exp}[ c N^{3/2}]$ and that there is a near perfect
cancellation between them in computing Euler characteristic,
and that there is a subleading term in the computation of Euler
characteristic that survives and that
would go as ${\rm exp}[a N]$.  This sounds on the face of it rather
like wishful thinking, but we will in fact prove that in the elliptic
case considered above this is exactly what happens!

We showed that the cohomologies of moduli space of holomorphic curves
are given by $n$-th level states in a fock space involving
bosons and fermions.  Note that if the number of bosons
and fermions were exactly equal we would get equal number
of even and odd cohomologies and thus zero Euler character.  In
fact if we wish to compute Euler character for the moduli
space wrapping the elliptic fiber $n$ times and the curve $C$ once,
instead of considering the coefficient of $q^n$ in
$${\prod_n(1+q^n)^{b^{odd}}\over \prod_n(1-q^n)^{b^{even}}}$$
we should weight the odd classes by a minus sign
and thus consider the coefficient of
$q^n$ in
$${\prod_n(1-q^n)^{b^{odd}}\over \prod_n(1-q^n)^{b^{even}}}
={1\over \prod_n(1-q^n)^{b^{even}-{b^{odd}}}}
$$
In our case $b^{even}-b^{odd}=12 c_1(C)$, which thus implies that
the coefficient of $q^n$ for the Euler character of moduli
spaces grows as
$${\rm exp}2\pi \sqrt{2 n c_1(C)}$$
Note that under uniform rescaling of charges by a factor of $N$
this grows as ${\rm exp}(aN)$ as expected on general grounds.
Having confirmed the general idea in this particular case
we can now be more confident and make the following prediction
(based on black hole analysis):  For large enough
degrees of curves, there will generically be a moduli space
of curves in CY 3-folds, with
nearly perfect equality of the number of even and odd cohomologies
where each goes as ${\rm exp}(b N^{3/2}) $ (where the coefficient of $b$
can be fixed using the black hole prediction as explained
above).  The subleading
difference in the growth of even and odd cohomologies should
then lead to a growth in the {\it Euler characteristic}
of moduli space as ${\rm exp}(a N)$.  It would be extremely
interesting to verify this prediction;  the case of
quintic 3-fold seems a good test case.

\newsec{Black holes in 4 dimensions}
We now consider type IIA strings compactified on a CY 3-fold
${\cal K}$ (not necessarily elliptic).  The black hole
entropy in this case has been recently computed in \msw\
using M-theory.
 Our derivation here uses type IIA to arrive at the same
conclusions, in the spirit of \gas \bsv .
 Consider a 4-cycle
realized through a holomorphic manifold ${\cal C}$.  Let us
assume $h^{1,0}({\cal C})=0$.  We are interested in a BPS
black hole which is a bound state of
$D4$ charge $[{\cal C}]$ and $n$ of $D0$
charge. According to \ref\astr{A. Strominger, Phys. Lett. {\bf B383}
(1996) 39.}\ref\klu{Behrndt et.al., ``Classical
and Quantum N=2 Supersymmetric Black Holes," hep-th/9610105.}\
 the entropy in this
case is predicted to be
\eqn\enen{S=2\pi \sqrt{ (n-{c_2({\cal C})\over 24})({1\over 6}{\cal C}^3)}}
where $c_2({\cal C})$ denotes the second chern class of
${\cal K}$ evaluated on ${\cal C}$ and ${\cal C}^3$ denotes the
triple self intersection of ${\cal C}$.

As was discussed in the previous section there are two ways to count
the microscopic entropy in this situation.  One using M-theory,
by considering M5 branes wrapped on ${\cal C}\times S^1$ with momentum
$n/R$ along $S^1$. This was the approach followed in \msw .  The
other is by considering the bound states of 0-brane and 4-brane, which
for $n$ large enough gives the answer (using the fact that since
$b^{odd}=0, b^{even}=\chi ({\cal C})$)
\eqn\bst{S=2\pi\sqrt{{1\over 6}n(\chi ({\cal C}))}}
 $\chi({\cal C})$ can be computed topologically in a standard
way \ket , using the fact
that the normal bundle of ${\cal C}$ is the canonical bundle
of ${\cal C}$, with the result
$$\chi({\cal C})={\cal C}^3+c_2({{\cal C}})$$
where $c_2({\cal C})$ denotes the second chern class of the CY 3-fold
evaluated on ${\cal C}$.
This is in agreement with the macroscopic entropy prediction \enen\
for $n>>C$, which is required for the use of Hardy-Ramanujan formula.
The microscopic term also includes a further linear correction in
${\cal C}$ which was interpreted in \msw\ as a one loop correction
for effective action in type IIA string of the form
$[\int_{\cal K} c_2\wedge k] R^2$ \bcov\
which gives a subleading correction to
the  black hole entropy.  As mentioned before this is similar
to what we have found in the 5 dimensional case as well.

We would like to thank S. Katz, J. Maldacena,
 V. Sadov, A. Strominger and S.-T. Yau for valuable discussions.

This research was supported in part by NSF grant PHY-92-18167.

\listrefs
\end